\documentclass[aps,prl,showpacs,twocolumn]{revtex4}
\usepackage{graphicx}
\usepackage{amsmath}%
\def\be{\begin{equation}}
\def\ee{\end{equation}}
\def\bea{\begin{eqnarray}}
\def\eea{\end{eqnarray}}

\def\A-1{$\AA^{-1}$}

\def\up{\uparrow}
\def\dn{\downarrow}
\def\~{$\approx$}

\begin{document}

\title{Superconductivity and magnetic order in 
CeRhIn$_{5}$; spectra of coexistence.}
\author{J.V. Alvarez and Felix Yndurain}
\affiliation{Departamento de F\'{\i}sica de la Materia Condensada,
Universidad Aut\'onoma de Madrid, 28049 Madrid, Spain}
\date{\today}
\begin{abstract}
We discuss the fixed-point Hamiltonian and the spectrum of excitations 
of a quasi-bidimensional electronic system supporting simultaneously  
antiferromamagnetic ordering and superconductivity. 
The coexistence of these two order parameters in a single phase 
is possible because the magnetic order is linked 
to the formation of a {\em metallic} spin density wave, 
and its order parameter is not associated to a spectral gap but 
to an energy shift of the paramagnetic bands. This peculiarity entails 
several distinct features in 
the phase diagram and the spectral properties of the model, which may have 
been observed in CeRhIn$_5$. Apart from the coexistence, we find
an abrupt suppression of the spin density wave
when the superconducting and magnetic ordering temperatures are equal. The divergence of the cyclotron mass extracted from de Haas-van Alphen experiments is also analyzed in the same framework. \\     
\end{abstract}
\maketitle
The interplay between magnetism and superconductivity is a recurrent area of
research in condensed matter physics. This interest has being activated 
in the last years due to the experimental findings of their coexistence 
in materials based on Ce, particularly in the 1-1-5 
CeMIn$_5$ family \cite{HEGGER00} \cite{NICKLAS04}. 
\begin{figure}[t]
\includegraphics[width=3.5in]{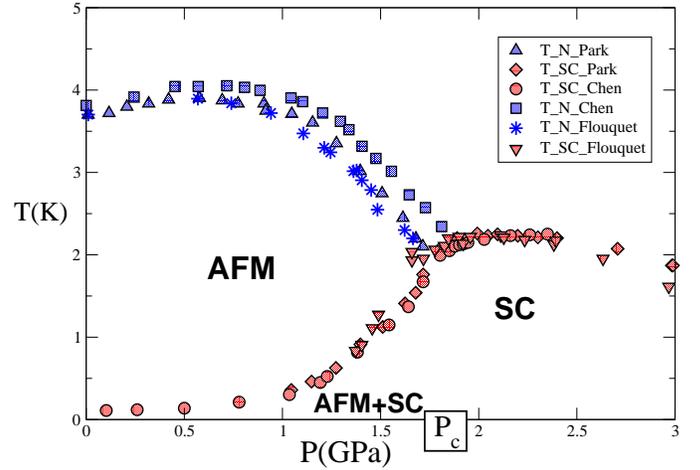}
\caption{Schematic experimental pressure-temperature phase diagram 
of CeRhIn$_5$ at zero-magnetic field adapted from 
references \cite{KNEBEL05}, \cite{CHEN06} and \cite{PARK07}.} 
\vspace{-0.1in} \label{exp_phd}
\end{figure}
A prominent member of this family is CeRhIn$_5$, 
which grows in tetragonal form, alternating CeIn$_3$  
and RhIn$_2$ planes along the c crystallographic axis. The structural 
anisotropy, induces quasi-bidimensionality in the electronic bonding 
and the Fermi surface, as evidenced in a series of de Haas-van Alphen 
measurements and band structure calculations 
\cite{dHvA-Band}\cite{SHISHIDO05dHvA}\cite{FUJIMORI}. 
At ambient pressure CeRhIn$_5$ becomes an antiferromagnet (AFM) 
below  $T_{\rm N}\sim 3.8$K \cite{HEGGER00}, with a  
small staggered magnetization aligned in the ab plane.   
Within the standard Doniach's Kondo lattice paradigm, 
applying pressure in a weak AFM heavy-fermion system opens a route to
very interesting effects. As the pressure increases 
this theoretical scenario predicts:     
i) A reduction of $T_N$ due to Kondo compensation.  
ii) The eventual suppression of the AFM order in a quantum critical 
point (QCP), which alike other heavy-fermion compounds, would be responsible 
of the anomalies  in the metallic phase.
iii) The  setting of unconventional superconductivity. 
However, all calorimetric \cite{KNEBEL05},  NQR \cite{MITO03NQR}, 
transport \cite{LLOBET04INS}  and susceptibility \cite{CHEN06} measurements 
provide a consistent picture for the pressure-temperature phase 
diagram (presented schematically in Fig. \ref{exp_phd}) in conflict 
with the aforementioned theoretical scenario.
Surprisingly, T$_N$ first increases with pressure and 
it only starts to decrease for 
pressures higher than 0.7GPa. Superconductivity shows up 
{\em before} $T_N$ has gone down to zero, 
i.e. AFM and SC {\em coexist}.  
Finally, AFM disappears {\em abruptly} at $P_c=1.9$GPa 
exactly when $T_N=T_{SC}$, in a first order transition and before 
a QCP could have taken place. 
Despite the lack of a QCP in the pressure-temperature phase diagram,  
the metallic phase still might be understood
in the framework of a quantum criticality if, changing another 
experimental knob, one could find a QCP nearby in the phase diagram. 
The natural choice is using a magnetic field 
to quench  the superconductivity, and in that way,  
continue the point $P_c(H=0)$ into a line of first order transitions 
down to zero temperature, ending with a QCP at $H_{c2}^{*}$. 
A new surprise appeared on this type of experiments \cite{KNEBEL05,PARK06}. 
For pressures higher than $P_c$, the AFM reenters applying an 
in-plane magnetic field $H_m<H_{c2}$.
Besides, transport measurements suggest \cite{HEGGER00,LLOBET04INS} 
that magnetic order in CeRhIn$_5$ may not be 
associated to a gap in the single-particle spectrum.
Actually, the resistivity as a function of the temperature 
does not show the conventional  minimum characteristic 
of a metal-insulator transition at any 
temperature. The small anomaly observed in the resistivity  
close to $T_N$ seems related to a change on the
scattering mechanism when the AFM sets in. 
 
In vivid contrast with quantum critical and quasi-one-dimensional systems, 
the understanding of the individual AFM or SC states rely on  
simple but accurate mean-field theories. 
In this Letter, we propose that a basic 
comprehension of the microscopic coexistence of AFM and SC 
and the first-order transition between these two phases can also 
be achieved within a mean-field scenario.
We discuss this phenomenon at the microscopic level in terms 
of a quasi-bidimensional model of interacting electrons proposed 
by one of us \cite{YNDURAIN91,YNDURAIN94,YNDURAIN94BOOK}. 
We will enumerate some implications of that model, discussing 
to which extent can be related to the phenomenology observed in 
CeRhIn$_{5}$.  

The model Hamiltonian contains the terms that naturally establish 
superconductivity and antiferromagnetism.  

\begin{equation}
H=H_{k}+H_{V}+H_{U}
\label{Hamiltonian}
\end{equation}
with
\begin{eqnarray}
H_{k}&=&\sum_{k \sigma} \varepsilon(k) c^{\dagger}_{k \sigma} c_{k \sigma}\\  
H_{V}&=&\sum_{kq} V_{kq} c^{\dagger}_{(k+q) \uparrow} c^{\dagger}_{(-k-q) \downarrow} c_{-k \downarrow} c_{k \uparrow} \\ 
H_{U}&=&\frac{U}{2} \sum_{kk'} c^{\dagger}_{k \sigma} c^{\dagger}_{k' -\sigma} c_{k  \sigma} c_{k' -\sigma}
\end{eqnarray}

Where U is a Hubbard on-site repulsion, and  $V_{kq}$  the 
effective interaction in the Cooper channel, which we will take 
to be attractive. Only s-wave pairing will be considered throughout this work.
Superconductivity with an order parameter having a symmetry related to  
$V_{kq}$ is favored by $H_{V}$ but the presence 
of a Hubbard repulsion establish a competition, which in terms of the gap    
is given by $\Delta=\Delta_2-\Delta_1$, such that:   
\begin{eqnarray}
\Delta_2&=&V \sum'_{k}  \langle c^{\dagger}_{k \sigma} c^{\dagger}_{-k -\sigma}
\rangle \\
\Delta_1&=&U \sum_{k}  \langle c^{\dagger}_{k \sigma} c^{\dagger}_{-k -\sigma}
\rangle 
\end{eqnarray}
where the prime restricts the summation to states with an energy 
(measured from the FL) smaller than a cut-off energy 
$E_c$. For those states we assume a very weak momentum dependence of 
$V_{kq}$.  In the absence of magnetic order and setting a constant density of states at the FL (i.e. far from a logarithmic divergence) , the competition between 
terms favoring and disfavoring the SC is clearly shown in the 
McMillan-like formula for the critical temperature.
\be
 T_{\rm SC}=1.13 E_c \exp \left(\frac{-1}{(V-U^{*})D(E_F)}\right)
\ee 
where $U^{*}=\frac{U}{1+U D(E_F)\ln(W/E_c)} $.

An essential ingredient of the model is the quasiparticle 
dispersion relation $\varepsilon(k)$ in the paramagnetic phase taken to set  
the Fermi level (FL) very close to the saddle point of a 2D system of 
itinerant electrons. Based on first principles calculations, Hall et al.\cite{dHvA-Band} find at the FL a sharp peak in the Density of States characteristic of a 2D van Hove logarithmic singularity. We have also performed a full first principles calculation of the electronic structure of tetragonal paramagnetic CeRhIn$_5$ using the VASP code \cite{VASP}. The generalized gradient approximation of Perdew et al.\cite{PERDEW} for the exchange and correlation was adopted. The results for the experimental lattice parameters \cite {MOSHOPOULOU} are reported in Figure 2. In Figure 2(a) we observe, like Hall et al.\cite{dHvA-Band}, a sharp peak in the density of states near the Fermi level. In addition, in Figure 2(b) we display the band structure at the vicinity of the Fermi level. We find a saddle point very close to the Fermi level whose dispersion along the z-direction (R-X-R in the standard notation) is about 0.3 eV indicating the two-dimensional character of the saddle point singularity.
\begin{figure}[t]
\includegraphics[bb=0 0 214 275,clip]{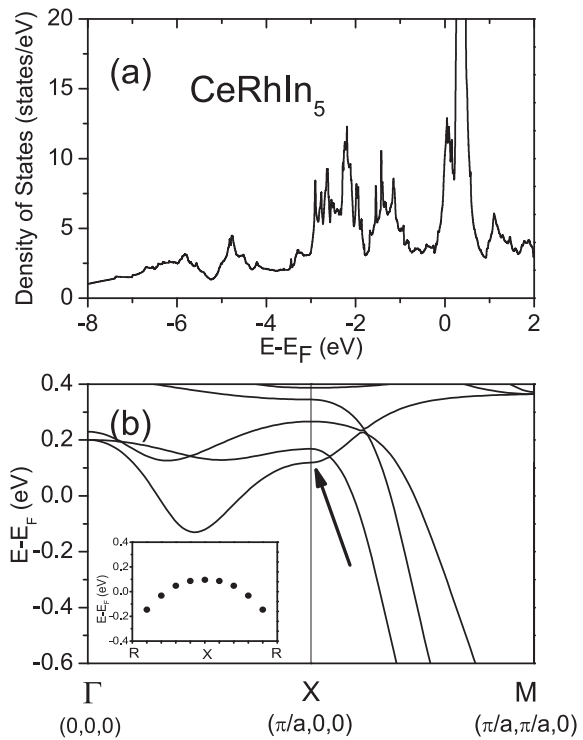}

\caption{ Calculated electronic structure of paramagnetic CeRhIn$_5$. (a) Density of states and (b) band structure. The arrow indicates a saddle point singularity at the X point of the Brillouin zone. The inset shows the variation of the saddle point energy along the z R-X-R direction (it crosses the Fermi Level at approximately the $(\pi/a,0,0.6\pi/c)$ point).}
\vspace{-0.1in} \label{vasp}
\end{figure}
Under these conditions we have, besides the Van Hove 
singularity in the density of states that favors superconductivity and magnetic order, an important kinematic restriction, 
namely $\varepsilon(k)=\varepsilon(k+Q)$, where $Q=(\pi/a,\pi/a)$ is the vector connecting two equivalent saddle points within the Brillouin zone. The above restriction determines the gapless nature 
of the SDW like in the CDW case proposed by Rice and Scott \cite{RICE75}.

The SDW order parameter is given by,
\be
\gamma_\sigma=U\sum_{k} \langle c^{\dagger}_{k \sigma} c_{k+Q -\sigma} \rangle
\ee
with $\gamma_{\sigma}=-\gamma_{-\sigma}=\gamma$. The resulting quadratic 
Hamiltonian is solved  by a Bogoliubov transformation and the SC and SDW order parameters obtained selfconsistently. The Hamiltionian eigenvalues are obtained by solving: 
\[ \left| \begin{array}{cccc}
\varepsilon_k-E_k & -\gamma_\up & -\Delta   & 0 \\
-\gamma_\up  & \varepsilon_{(k+Q)}-E_k & 0 & -\Delta \\
-\Delta & 0 & \varepsilon_{-k}-E_k & \gamma_\dn \\
0 & -\Delta & \gamma_\dn & \varepsilon_{-(k+Q)}-E_k  \\
\end{array} \right|=0 \] 
The four solutions are $E_1=-E_{-},E_2=-E_{+},E_3=E_{-},E_4=E_{+}$
where $E_{\pm}(k)=\sqrt{ \varepsilon(k)^2+\Delta^2}\pm \gamma$. Notice that for a 1D or 2D nested system $\varepsilon(k)=-\varepsilon(k+Q)$ and therefore $E_{\pm}(k)=\pm \sqrt{ \varepsilon(k)^2+\Delta^2 +\gamma^2}$. 

\begin{figure}[t]
\includegraphics[width=3.2in]{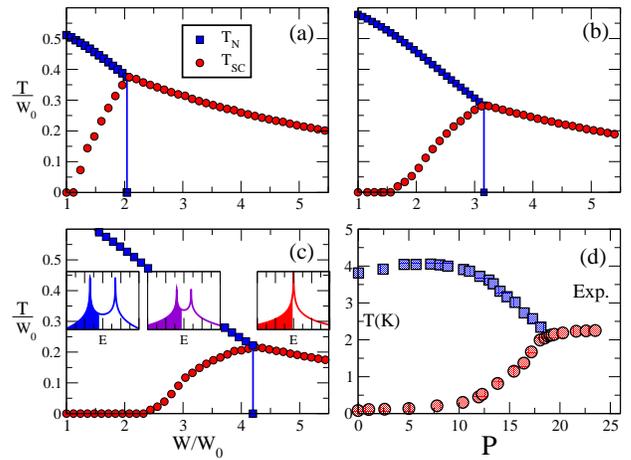}

\caption{ Phase diagram (temperature  versus pressure) obtained using the 
model Hamiltonian (\ref{Hamiltonian}) and a two dimensional band structure 
with the FL close to a saddle point. The variables are in units of the half bandwidth $W_0$.  $T_N$ and $T_{SC}$ stand for the SDW and SC critical temperatures respectively. Panels (a), (b) and (c) are the calculated results for the parameters given in the text and $U=2.25W_{0}$ , $U=2.50W_{0}$ and $U=2.75W_{0}$ respectively. The inset in panel (c) indicates the metallic 
DOS in the three (SDW, SDW+SC and SC) different regimes. The shaded area
indicates occupied levels. Panel (d) represents the experimental results of ref.  \cite{CHEN06}}
\vspace{-0.1in} \label{th_phd}
\end{figure}

The system is solved self-consistently for an effective half bandwidth
at ambient pressure of $W_0$ and the phase diagram 
for $V=4W_{0}$, $E_c=0.7W_{0}$ , $n=0.92$ electrons and $U=2.25W_{0}$ , $U=2.50W_{0}$ and $U=2.75W_{0}$, 
is presented in Fig. \ref {th_phd}. To simulate the effect of the pressure we 
have considered a linear variation of the bandwidth with the pressure and
a bandwidth independent electron-electron interaction U.
These assumptions have being found to be reasonable with a first principles calculation 
using the SIESTA code \cite{SIESTA} taking Ni as a benchmark. 
Also, for simplicity, a constant V interaction and hole concentration independent of pressure are assumed. 
The results of our model agree qualitatively with the experimental 
findings depicted in Fig. \ref{exp_phd}and Fig. 3(d). We indeed find a competition between SDW and SC but, 
as seen experimentally, 
they can coexist in a non negligible region of the phase diagram. In addition, 
the SDW disappears abruptly when the two critical temperatures became equal, i.e., the SDW transition temperature can not be lower than the superconducting one. This numerical result is similar to the analytical finding by Bilbro and McMillan \cite{BILBRO} concerning superconductivity and martensitic transformation in A15 compounds. 

\begin{figure}[t]
\includegraphics[width=3.2in]{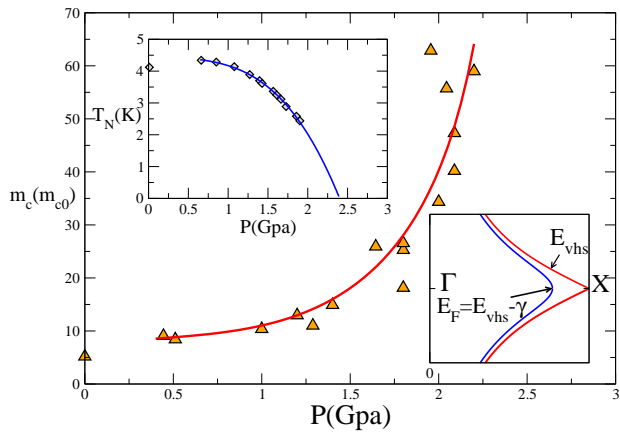}
\caption{Cyclotron mass ($m_c$) as a function of pressure. 
Triangles are the experimental values from Ref. \cite{SHISHIDO05dHvA} 
showing the $m_c$ divergence close to $P \sim 2.4$ GPa, where the 
magnetic order disappears. The solid line is a fit to the  theoretical model.  
Close to a Van Hove singularity the cyclotron mass diverges 
logarithmically as the difference between $E_F$ and the $E_{vhs}$ vanishes
(see lower inset).   
According to the model this energy difference 
is proportional to $T_N$ and its pressure dependence can be extracted, 
for instance, from Ref. \cite{KNEBEL05}(see upper inset).}
\vspace{-0.1in} \label{m_c_P}
\end{figure}

The proximity of the FL to a saddle point is an ingredient of model (\ref{Hamiltonian}) necessary  
for the formation of the metallic SDW. 
We ask ourselves whether in CeRhIn$_5$ there is experimental evidence for such a proximity.    
A recent de Haas-van Alphen study \cite{SHISHIDO05dHvA} shows a 
divergence in the cyclotron mass at a pressure $P_{c2} \sim 2.4$GPa, 
accompanied with a change in the quasi-2D Fermi surface. 
Those experiments were performed for values of magnetic fields and pressures 
in which the system is AFM.  Following Park et {\em al.}  $P_{c2}$ 
is very close to the pressure at which $T_N$ would extrapolate to zero
in absence of SC (see upper inset in Fig. \ref{m_c_P} ). 
The cyclotron mass in a 2D electronic system 
$m_c=\hbar^2/2\pi(\partial A(E)/\partial E )$ where A is the area enclosed 
in the isoenergetic contour line 
$E(k)=E$. Close to the 2D Van Hove singularity one 
expects the cyclotron mass to diverge logarithmically. Actually, the 
precise functional form has been computed  in Ref. \cite{ITSKOVSKY05dHvH_VHS} 
and found to be   
\be
m_c=m_{c0}\left(C+D\ln{\frac{E_{vhs}}{|E_F-E_{vhs}|}}\right)
\label{m_c_fit}
\ee
In a tetragonal crystal structure C and D are numbers nearly independent 
of the pressure and $m_{c0}$  is the cyclotron mass at the bottom of the band. 
The divergence is driven by the denominator in the argument of the logarithm 
which in our model is: $E_F(\gamma)-E_{vhs} \sim \gamma(P) \sim T_N(P)$. 
In other words, in model (\ref{Hamiltonian}),  $m_c$ is enhanced  as the 
AFM vanishes because the FL in the SDW  approaches the saddle point at X 
(see the inset in Fig. \ref{m_c_P}).
To elucidate if the experimental results are compatible 
with this argument and with expression (\ref{m_c_fit}), 
we have extracted the pressure dependence of $T_N$, fitting  the experimental data in Ref (\cite{KNEBEL05}) (see upper inset in Fig. \ref{m_c_P} to a cubic 
polynomial law in $(P_{c2}-P)$ in the range of pressures
between $P=0.65$ and $P=2.4$ GPa. We do not attempt to justify physically 
this fit here,  since our only goal is extracting an analytic expression 
for $T_N(P)$ for the range of pressures of interest, to insert it in 
(\ref{m_c_fit}). The results are presented in the main panel of 
Fig. \ref{m_c_P}. Our model reproduces reasonably well 
these experimental findings.

Within this model, we expect an anomaly in the specific heat at $T_N$, 
which is not associated to a SDW spectral gap but to the  
entropy released when the magnetic order disappears. 
The electronic part of the specific heat is: 
\begin{eqnarray}
C^{e-}_{V}=-2 \beta \sum_{i,k} E_{i}(k) \frac{\partial f_k}{\partial E_{i}(k)}
\times \\
\left(E_{i}(k)+\frac{\beta}{2 \sqrt{\varepsilon(k)^2+\Delta^2}}\frac{d \Delta^2}{d \beta}+\beta \frac{d \gamma}{d \beta}\right)
\end{eqnarray}
where i=3,4. The second term inside the parenthesis gives the SC anomaly at 
$T_C$  and the third term gives the antiferromagnetic anomaly at $T_N$.
The SC anomaly is much weaker in the coexisting phase,  
because the FL  
lies in a depression of the density of states 
(see central graph on the panel (c) of Fig. \ref{th_phd}) 
created by the underlying SDW,  while in the purely 
SC phase the FL is very close to a divergence in the density of states (right graph). 
Remarkably, this enhancement 
has been also observed in calorimetric measurements on CeRhIn$_5$ 
\cite{KNEBEL05}

To summarize: beyond the detailed boundary shape in the 
phase diagram of CeRhIn$_5$ we have identified two unequivocal, 
clear-cut features of the phenomenology of CeRhIn$_5$ in the discussed model, 
namely; coexistence and abrupt disappearance of 
AFM when $T_N=T_{SC}$.
The essential ingredient in the model is the metallic SDW, favored for the proximity of Fermi level to a Van Hove logarithmic singularity in the density of states.
The gapless nature of the SDW implies
the lack of a metal-insulator transition at, or close to, $T_N$
as shown by resistivity measurements. The kinematic conditions needed for 
the metallic SDW to be formed seem to be present 
in CeRhIn$_5$ as shown by the logarithmic divergence of the 
cyclotron mass.

We  appreciate very much discussions with G. Gomez-Santos, H. Suderow and  
S. Vieira. Financial support of
the Spanish Ministry of Science ( Ramon y Cajal 
contract and Grant BFM2003-03372) is acknowledged.

\end{document}